\documentclass[12pt]{article}
\usepackage{amsmath}
\usepackage{amssymb}

\newcommand{\pt}{$\cal PT$}
\newcommand{\ap}{a^{\dagger}}
\newcommand{\R}{\mathbb{R}}
\newcommand{\C}{\mathbb{C}}
\DeclareMathOperator{\arctanh}{arctanh}
\newcommand{\nablab}{\mbox{\boldmath $\nabla$}}
\newcommand{\xb}{\mbox{\boldmath $x$}}

\title{
Non-Hermitian oscillator Hamiltonian and su(1,1): a way towards generalizations}
\author{C Quesne\\ 
{\small Physique Nucl\'eaire Th\'eorique et Physique Math\'ematique,  Universit\'e Libre de Bruxelles,} \\ 
{\small Campus de la Plaine CP229, Boulevard~du Triomphe, B-1050 Brussels, Belgium}\\
{\small E-mail: cquesne@ulb.ac.be}}
\date{ }
\begin{document}
\baselineskip=22pt plus 1pt minus 1pt
%%%%%%%%%%%%%%%%%%%%%%%%%%%%%%%%%%%%%%%%%%%%%%%%%%%%%%%%%%
\maketitle

\begin{abstract} 
The family of metric operators, constructed by Musumbu {\sl et al} (2007 {\sl J.\ Phys.\ A: Math.\ Theor.} {\bf 40} F75), for a harmonic oscillator Hamiltonian augmented by a non-Hermitian $\cal PT$-symmetric part, is re-examined in the light of an su(1,1) approach. An alternative derivation, only relying on properties of su(1,1) generators, is proposed. Being independent of the realization considered for the latter, it opens the way towards the construction of generalized non-Hermitian (not necessarily $\cal PT$-symmetric) oscillator Hamiltonians related by similarity to Hermitian ones. Some examples of them are reviewed.
\end{abstract}

\noindent
Short title: Non-Hermitian oscillator Hamiltonian and su(1,1)


%
%========================================================================
%
\newpage
Non-Hermitian Hamiltonians $H$ with a real spectrum are currently an active field of research, motivated both by the necessity to understand their mathematical properties and by the requirement to build a consistent unitary theory of quantum mechanics for them (see \cite{jpa, czech} for recent surveys).\par
%
%----------------------------------------------------------------------------------------------------------
%
It has been shown that the latter demands the existence in the relevant Hilbert space $\cal H$ of a positive-definite metric operator $\zeta_+$, defining a new inner product $\langle \cdot, \cdot \rangle_+ = \langle \cdot, \zeta_+ \cdot \rangle$, with respect to which $H$ becomes Hermitian. Such a Hamiltonian then possesses a Hermitian counterpart $h = \rho H \rho^{-1}$ (where $\rho = \sqrt{\zeta_+}$) with respect to the original inner product $\langle \cdot, \cdot \rangle$ and is termed quasi-Hermitian \cite{scholtz92}.\par
%
%---------------------------------------------------------------------------------------------------------
% 
Quasi-Hermitian Hamiltonians can be seen as a subclass of pseudo-Hermitian ones, for which there exists a Hermitian invertible (not necessarily positive-definite) operator $\zeta$ such that $H^{\dagger} = \zeta H \zeta^{-1}$ \cite{mosta02, mosta04}. In particular, for \pt-symmetric (or $\cal P$-pseudo-Hermitian) Hamiltonians \cite{bender98}, the metric operator $\zeta_+$ is related to the so-called charge operator $\cal C$ and $\langle \cdot, \cdot \rangle_+$ is also called $\cal CPT$-inner product \cite{bender02, bender04}.\par
%
%---------------------------------------------------------------------------------------------------
%  
The metric operator $\zeta_+$ gives rise to two problems. First, as it is dynamically determined, it must be constructed for every non-Hermitian Hamiltonian that is suspected to be quasi-Hermitian. Second, as it is not unique, one may wonder what is the physical significance of a specific choice. It has been argued \cite{scholtz92} that selecting a given $\zeta_+$ amounts to choosing which additional operators are quasi-Hermitian with respect to the same metric as $H$ and form with the latter an irreducible set of observables on $\cal H$.\par
%
%-----------------------------------------------------------------------------------------------------
% 
This last point has been recently illustrated \cite{musumbu} with a simple non-Hermitian \pt-symmetric oscillator Hamiltonian, first proposed by Swanson \cite{swanson} and later on studied by various authors \cite{geyer, jones, bagchi, scholtz06a, scholtz06b}. This Hamiltonian is given by
\begin{equation}
  H = \omega \bigl(\ap a + \tfrac{1}{2}\bigr) + \alpha a^2 + \beta a^{\dagger2}  \label{eq:swanson}
\end{equation}
where $\ap = (\omega x - {\rm i} p)/\sqrt{2\omega}$ and $a = \bigl(\ap\bigr)^{\dagger}$ are standard harmonic oscillator creation and annihilation operators (with $p = - {\rm i} d/dx$ and $\hbar = m = 1$), while $\omega$, $\alpha$, $\beta$ are three real parameters such that $\alpha \ne \beta$ and $\omega^2 - 4 \alpha \beta > 0$. A family of metric operators $\zeta_+$ (denoted by $\Theta = S^2$ in \cite{musumbu}), depending on a continuous variable $z \in [-1, 1]$, has been constructed by a generalized Bogoliubov transformation approach similar to that introduced by Swanson. The corresponding Hermitian Hamiltonian $h$ and the quasi-Hermitian position and momentum operators have also been built in terms of $z$. In the generic case, such quasi-Hermitian operators are non-Hermitian (in the original $L^2$ sense) linear combinations of $x$ and $p$. There exists a quadratic function of them, denoted by $O$, which is Hermitian with respect to the original inner product $\langle \cdot, \cdot \rangle$. An irreducible set of observables is therefore made of $H$ and $O$ or, in the equivalent Hermitian description, $h$ and $O$.\par
%
%--------------------------------------------------------------------------------------------------
% 
The purpose of the present paper is to propose another type of approach than that considered in \cite{musumbu, swanson}. It is based on the fact that the Hamiltonian (\ref{eq:swanson}) can be written as a linear combination of su(1,1) generators $K_0$, $K_+$, $K_-$,
\begin{equation}
  H = 2\omega K_0 + 2\alpha K_- + 2\beta K_+.  \label{eq:gen-Swanson}
\end{equation}
Here
\begin{equation}
  K_0 = \tfrac{1}{2} \bigl(\ap a + \tfrac{1}{2}\bigr) \qquad K_+ = \tfrac{1}{2} a^{\dagger2} \qquad
  K_- = \tfrac{1}{2} a^2  \label{eq:HO}
\end{equation}
indeed satisfy the defining su(1,1) commutation relations
\begin{equation}
  [K_0, K_{\pm}] = \pm K_{\pm} \qquad [K_+, K_-] = - 2 K_0  \label{eq:com}
\end{equation}
and Hermiticity properties
\begin{equation}
  K_0^{\dagger} = K_0 \qquad K_{\pm}^{\dagger} = K_{\mp}.  \label{eq:adjoint}
\end{equation}
We plan to show that for those Hamiltonians that can be expressed as (\ref{eq:gen-Swanson}), equations (\ref{eq:com}) and (\ref{eq:adjoint}) yield a family of metric operators $\zeta_+ = \rho^2$, which in the case of (\ref{eq:swanson}), i.e., for realization (\ref{eq:HO}), reduces to that considered in \cite{musumbu}. Since, however, the validity of our new derivation is independent of the chosen realization, it applies to other Hamiltonians satisfying equation (\ref{eq:gen-Swanson}). In the second part of this paper, we will provide some examples of such Hamiltonians, which may be called `generalized non-Hermitian oscillator Hamiltonians'.\par
%
%------------------------------------------------------------------------------------------------------
%
Let us start with an ansatz for the Hermitian operator $\rho$ similar to that used in \cite{musumbu}, namely
\begin{equation}
  \rho = \exp(A) \qquad A = 2\epsilon K_0 + 2\eta K_- + 2\eta^* K_+  \label{eq:rho}
\end{equation}
where we assume $\epsilon \in \R$, $\eta \in \C$ and $\theta^2 = \epsilon^2 - 4 |\eta|^2 \ge 0$ (meaning that $\theta \in \R$), and let us determine under which conditions on $\epsilon$ and $\eta$ the operator $h = \rho H \rho^{-1}$ is Hermitian.\par
%
%--------------------------------------------------------------------------------------------------
% 
Since $H$ is a linear combination of $K_0$, $K_+$ and $K_-$, the calculation of the action of $\rho$ on it amounts to that of $\rho$ on the generators. To this end, it is useful to factorize $\rho$, defined in (\ref{eq:rho}), in either of the following two forms
\begin{equation}
  \rho = \exp(p K_+) \exp(q K_0) \exp (r K_-) = \exp(r' K_-) \exp(q' K_0) \exp(p' K_+)  \label{eq:factor}
\end{equation}
for some parameters $p$, $q$, $r$, $p'$, $q'$, $r' \in \C$. Such factorizations can actually be applied to the exponential of any linear combination of $K_0$, $K_+$ and $K_-$ with complex coefficients because they only result from the commutation relations (\ref{eq:com}), independently of the Hermiticity conditions chosen for the generators. In other words, they are sl(2) properties, which means that the parameter values can be determined \cite{gilmore, truax} by realizing equation (\ref{eq:factor}) in some faithful sl(2) representation, e.g., the $2 \times 2$ matrix one
\begin{equation}
  \sigma(K_0) = \frac{1}{2} \begin{pmatrix} 1 & 0 \\ 0 & -1 \end{pmatrix} \qquad \sigma(K_+) =
  \begin{pmatrix} 0 & 1 \\ 0 & 0 \end{pmatrix} \qquad \sigma(K_-) = \begin{pmatrix} 0 & 0 \\ -1 & 0
  \end{pmatrix}.  \label{eq:sigma}
\end{equation}
A straightforward calculation then leads to the results
\begin{align*}
  e^{-q/2} &= \cosh \theta - \epsilon \frac{\sinh \theta}{\theta} & r &= p^* = \frac{2\eta \sinh \theta/
    \theta}{\cosh \theta - \epsilon \sinh \theta/\theta} \\
  e^{-q'/2} &= \cosh \theta + \epsilon \frac{\sinh \theta}{\theta} & r' &= p^{\prime*} = \frac{2\eta 
    \sinh \theta/\theta}{\cosh \theta + \epsilon \sinh \theta/\theta}
\end{align*}
where we can check that for the special case of a Hermitian $\rho$, which we consider here, the factorized operators are also Hermitian, as it should be.\par
%
%-----------------------------------------------------------------------------------------------------
%
On using the Baker-Campbell-Hausdorff formula or the matrix representation (\ref{eq:sigma}), it is now an easy task to prove the relations
\begin{align*}
  \rho K_0 \rho^{-1} &= \left(1 - 8 |\eta|^2 \frac{\sinh^2 \theta}{\theta^2}\right) K_0 + 2 \eta
    \frac{\sinh \theta}{\theta} \left(\cosh \theta - \epsilon \frac{\sinh \theta}{\theta}\right) K_- \\
  & \quad - 2 \eta^* \frac{\sinh \theta}{\theta} \left(\cosh \theta + \epsilon \frac{\sinh \theta}{\theta}  
    \right) K_+ \\
  \rho K_- \rho^{-1} &= - 4 \eta^* \frac{\sinh \theta}{\theta} \left(\cosh \theta - \epsilon \frac{\sinh 
    \theta}{\theta}\right) K_0 + \left(\cosh \theta - \epsilon \frac{\sinh \theta}{\theta}\right)^2 K_-
    + 4 \eta^{*2} \frac{\sinh^2 \theta}{\theta^2} K_+ \\
  \rho K_+ \rho^{-1} &= 4 \eta \frac{\sinh \theta}{\theta} \left(\cosh \theta + \epsilon \frac{\sinh 
    \theta}{\theta}\right) K_0 + 4 \eta^2 \frac{\sinh^2 \theta}{\theta^2} K_- + \left(\cosh \theta +
    \epsilon \frac{\sinh \theta}{\theta}\right)^2 K_+.    
\end{align*}
\par
%
%----------------------------------------------------------------------------------------------------
%
On combining these results with definition (\ref{eq:gen-Swanson}), we obtain
\begin{equation*}
  \rho H \rho^{-1} = 2 U K_0 + 2 V K_- + 2 W K_+
\end{equation*}
where
\begin{align*}
  U &= \omega \left(1 - 8 |\eta|^2 \frac{\sinh^2 \theta}{\theta^2}\right) - 4 \alpha \eta^* \frac{\sinh 
    \theta}{\theta} \left(\cosh \theta - \epsilon \frac{\sinh \theta}{\theta}\right) \\
  & \quad + 4 \beta \eta \frac{\sinh \theta}{\theta} \left(\cosh \theta + \epsilon \frac{\sinh \theta}
    {\theta}\right) \\
  V &= 2 \omega \eta \frac{\sinh \theta}{\theta} \left(\cosh \theta - \epsilon \frac{\sinh \theta}{\theta}
    \right) + \alpha \left(\cosh \theta - \epsilon \frac{\sinh \theta}{\theta}\right)^2 + 4 \beta \eta^2
    \frac{\sinh^2 \theta}{\theta^2} \\
  W &= - 2 \omega \eta^* \frac{\sinh \theta}{\theta} \left(\cosh \theta + \epsilon \frac{\sinh \theta}
    {\theta}\right) + 4 \alpha \eta^{*2} \frac{\sinh^2 \theta}{\theta^2} + \beta \left(\cosh \theta + 
    \epsilon \frac{\sinh \theta}{\theta}\right)^2 
\end{align*}
coincide with similar quantities considered in equation (8) of \cite{musumbu}. We may therefore directly assert that the Hermiticity conditions of $h = \rho H \rho^{-1}$, i.e., $U^* = U$ and $W^* = V$, read
\begin{equation}
  \eta^* = \eta \qquad \frac{\tanh 2\theta}{\theta} = \frac{\alpha - \beta}{(\alpha + \beta) \epsilon
  - 2 \omega \eta}  \label{eq:conditions}
\end{equation}
and that the second condition in (\ref{eq:conditions}) determines $\epsilon$ as
\begin{equation*}
  \epsilon = \frac{1}{2\sqrt{1 - z^2}} \arctanh \frac{(\alpha - \beta) \sqrt{1 - z^2}}{\alpha + \beta -
  z \omega}
\end{equation*}
in terms of the parameter $z = 2\eta/\epsilon$, where $-1 \le z \le 1$.\par
%
%----------------------------------------------------------------------------------------------------
%
On transposing corresponding results of \cite{musumbu}, we may write the Hermitian counterpart of $H$, the transformation $\rho$ and the additional observable $O$ (such that $[\rho, O] = 0$) as
\begin{align}
  h &= \frac{1}{2\omega} \bigl[\nu (2K_0 + K_+ + K_-) + \mu \omega^2 (2K_0 - K_+ - K_-)\bigr] 
    \label{eq:h}  \\
  \rho &= \left(\frac{\alpha + \beta - \omega z + (\alpha - \beta) \sqrt{1 - z^2}}{\alpha + \beta - \omega 
    z - (\alpha - \beta) \sqrt{1 - z^2}}\right)^{[2K_0 + z (K_+ + K_-)]/(4 \sqrt{1 - z^2})} \\
  O &= 2K_0 + z (K_+ + K_-) 
\end{align}
respectively. In (\ref{eq:h}), $\mu$ and $\nu$ are defined by
\begin{align*}
  \mu &= [(1+z) \omega]^{-1} \left[\omega - (\alpha + \beta) z - (\alpha + \beta - \omega z) \left(1 -
  \frac{(\alpha - \beta)^2 (1 - z^2)}{(\alpha + \beta - \omega z)^2}\right)^{1/2} \right] \\
  \nu &= (1-z)^{-1} \omega \left[\omega - (\alpha + \beta) z + (\alpha + \beta - \omega z) \left(1 -
  \frac{(\alpha - \beta)^2 (1 - z^2)}{(\alpha + \beta - \omega z)^2}\right)^{1/2} \right].
\end{align*}
This completes the extension of \cite{musumbu} to the whole family of generalized non-Hermitian oscillator Hamiltonians, defined in (\ref{eq:gen-Swanson}).\par
%
%--------------------------------------------------------------------------------------------------------------------
%
Let us now review some examples of such Hamiltonians. This amounts to considering various physically-relevant realizations of the su(1,1) generators.\par
%
%-------------------------------------------------------------------------------------------------------------------
%
%
To start with, we may consider a straightforward generalization dealing with a non-Hermitian $d$-dimensional radial harmonic oscillator
\begin{equation*}
  H = \frac{1}{2\omega} \left[(\omega - \alpha - \beta)\left(- \frac{d^2}{dr^2} + \frac{L(L+1)}{r^2}
  \right) + (\omega + \alpha + \beta) \omega^2 r^2 + (\alpha - \beta) \omega \left(2r \frac{d}{dr} + 1
  \right)\right]
\end{equation*}
constructed from the operators
\begin{align*}
  K_0 &= \frac{1}{4\omega} \left(- \frac{d^2}{dr^2} + \frac{L(L+1)}{r^2} + \omega^2 r^2\right) \\
  K_{\pm} &= \frac{1}{4\omega} \left[\frac{d^2}{dr^2} - \frac{L(L+1)}{r^2} + \omega^2 r^2 \mp \omega 
    \left(2r \frac{d}{dr} + 1\right)\right].   
\end{align*}
Here $r$ runs over the half-line $0 < r < \infty$ and $L$ is defined by $L = l + (d-3)/2$ in terms of the angular momentum quantum number $l$. In such a case, since the angular variables remain unaffected by the Hermiticity breaking, an irreducible set of observables can be obtained by supplementing $h$ and $O$ with some standard angular ones.\par
%
%--------------------------------------------------------------------------------------------------
%
Another rather direct extension consists in choosing a more general realization \cite{bagchi}
\begin{equation*}
  a = A(x) \frac{d}{dx} + B(x) \qquad \ap = - A(x) \frac{d}{dx} + B(x) - A'(x) \qquad 2AB' - AA'' = 1
\end{equation*}
for the annihilation and creation bosonic operators appearing in the original Hamiltonian (\ref{eq:swanson}). Here a prime denotes derivative with respect to $x$. On setting
\begin{equation*}
  g(x) = \int^x \frac{dx'}{A(x')} \qquad B(x) = - \frac{g''}{2g^{\prime2}} + \frac{1}{2} g + \tau
\end{equation*}
where $\tau$ is some integration constant, the generator realization becomes
\begin{align*}
  K_0 &= \frac{1}{2}\left[- \frac{d}{dx} \frac{1}{g^{\prime2}} \frac{d}{dx} + \frac{g'''}{2g^{\prime3}} -
    \frac{5}{4} \frac{g^{\prime\prime2}}{g^{\prime4}} + \left(\frac{1}{2} g + \tau\right)^2\right] \\
  K_{\pm} &= \frac{1}{2}\left[\frac{d}{dx} \frac{1}{g^{\prime2}} \frac{d}{dx} \mp \frac{1}{g'} (g + 2\tau)
    \frac{d}{dx} - \frac{g'''}{2g^{\prime3}} + \frac{5}{4} \frac{g^{\prime\prime2}}{g^{\prime4}} \pm
    \frac{g''}{g^{\prime2}} \left(\frac{1}{2} g + \tau\right) + \left(\frac{1}{2} g + \tau\right)^2 \mp
    \frac{1}{2}\right]  
\end{align*}
in terms of some function $g(x)$. It is then clear that the resulting Hamiltonian (\ref{eq:gen-Swanson}) is a non-Hermitian position-dependent mass (PDM) Hamiltonian with a mass proportional to $g^{\prime2}$ and that it is not \pt\ symmetric unless $g(x)$ is an even function. For the choice $g(x) = - e^{-px}/p$ ($p \in \R$), for instance, we get a non-Hermitian Hamiltonian equivalent to PDM Hermitian ones of the form
\begin{equation}
  h = - \frac{1}{2} \frac{d}{dx} \frac{1}{m(x)} \frac{d}{dx} + V_{\rm eff}(x)  \label{eq:h-PDM}
\end{equation}
with an exponential mass and a Morse-like potential
\begin{align*}
  m(x) &= \frac{1}{2\mu\omega} e^{-2px} \\
  V_{\rm eff}(x) &= - \frac{3}{4} \mu \omega p^2 e^{2px} + \frac{\nu}{\omega} \left(- \frac{1}{2p}
    e^{-px} + \tau\right)^2.
\end{align*}
Note that the special case of (\ref{eq:h-PDM}) corresponding to $z=1$ has been given in \cite{bagchi}.\par
%
%--------------------------------------------------------------------------------------------------------------
%
A further example with applications in quantum optics is provided by multiboson realizations of su(1,1) (see \cite{golinski} and references quoted therein). For one-mode systems, for instance, we have
\begin{equation*}
  K_0 = \alpha_0(N) \qquad K_- = \alpha_-(N) a^l \qquad K_+ = a^{\dagger l} \alpha_-(N)
\end{equation*}
where $N = \ap a$, $l$ is some fixed positive integer and $\alpha_0(N)$, $\alpha_-(N)$ are some real functions of $N$, which can be expressed as
\begin{align*}
  \alpha_0(N) &= \frac{1}{l}(N - R) + \alpha_0(R) \\
  \alpha_-(N) &= \left[\frac{1}{(N+1)_l} \left(\frac{1}{l}(N - R) + 2\alpha_0(R)\right) \left(\frac{1}{l}(N - R) 
  + 1\right)\right]^{1/2}.
\end{align*}
Here $(N+1)_l = (N+1) (N+2) \dotsm (N+l)$ and
\begin{equation*}
  R = \begin{cases}
    0& \text{if $l=1$} \\[0.2cm]
    \frac{l-1}{2} + \sum_{m=1}^{l-1} \frac{\exp(- 2\pi {\rm i} mN/l)}{\exp(2\pi {\rm i} m/l) - 1}& 
      \text{if $l>1$}  
  \end{cases}
\end{equation*}
acts on $n$-boson states $|n\rangle = (n!)^{-1/2} a^{\dagger n} |0\rangle$ as $R |n\rangle = n\bmod l\, |n\rangle$. For such a realization, Hamiltonians of type (\ref{eq:gen-Swanson}) with $\alpha = \beta$ are currently employed to describe parametric absorption-emission of one-mode bosons in nonlinear media. The present work therefore enables us to extend such applications to non-Hermitian Hamiltonians with $\alpha \ne \beta$ since we have shown that they have a Hermitian counterpart $h$ of the usual form.\par
%
%----------------------------------------------------------------------------------------------------------
% 
The last example we would like to mention here embraces the set of conformal $d$-dimensional $n$-body systems (see \cite{meljanac05} and references quoted therein), described by Hamiltonians of the type
\begin{equation}
  H = - \frac{1}{2} \sum_{i=1}^n \frac{1}{m_i} \nablab_i^2 + V(\xb_1, \ldots, \xb_n) + \frac{1}{2}
  \omega^2 \sum_{i=1}^n m_i \xb_i^2 + \frac{1}{2} c \left(\sum_{i=1}^n \xb_i \cdot \nablab_i + 
  \frac{1}{2} nd\right).  \label{eq:conformal}  
\end{equation}
Here $m_i$, $i=1$, 2, \ldots, $n$, are the particle masses, $x_{i\alpha}$, $i=1$, 2, \ldots, $n$, $\alpha = 1$, 2, \ldots, $d$, their coordinates, $c$ is some real constant and $V(\xb_1, \ldots, \xb_n)$ is a real\footnote{It should be noted that in \cite{meljanac05}, $V(\xb_1, \ldots, \xb_n)$ may be real or \pt\ symmetric. In the latter case, however, the procedure considered here would not work.} function of order $-2$, i.e., such that $[\xb_i \cdot \nablab_i, V] = - 2V$. The corresponding realization of the su(1,1) generators reads
\begin{align*}
  K_0 &= \frac{1}{2\omega} \left(- \frac{1}{2} \sum_{i=1}^n \frac{1}{m_i} \nablab_i^2 + V(\xb_1, \ldots, 
    \xb_n) + \frac{1}{2} \omega^2 \sum_{i=1}^n m_i \xb_i^2\right) \\
  K_{\pm} &= \frac{1}{2\omega} \left[\frac{1}{2} \sum_{i=1}^n \frac{1}{m_i} \nablab_i^2 - V(\xb_1, 
    \ldots, \xb_n) + \frac{1}{2} \omega^2 \sum_{i=1}^n m_i \xb_i^2 \mp \omega \left(\sum_{i=1}^n \xb_i 
    \cdot \nablab_i + \frac{1}{2} nd\right)\right]
\end{align*}  
and their combination (\ref{eq:gen-Swanson}) coincides with Hamiltonian (\ref{eq:conformal}) provided we choose $\alpha = - \beta = c/4$. The family of equivalent Hermitian Hamiltonians is then given by
\begin{equation}
  h = \mu \left(- \frac{1}{2} \sum_{i=1}^n \frac{1}{m_i} \nablab_i^2 + V(\xb_1, \ldots, \xb_n)\right) + 
  \frac{1}{2} \nu \sum_{i=1}^n m_i \xb_i^2  \label{eq:h-conformal} 
\end{equation}
where
\begin{equation*}
  \mu = (1+z)^{-1} \left[1 + \frac{z}{\omega |z|} \left(\Omega^2 z^2 - \frac{c^2}{4}\right)^{1/2}\right]
  \qquad \nu = (1-z)^{-1} \omega^2 \left[1 - \frac{z}{\omega |z|} \left(\Omega^2 z^2 - \frac{c^2}{4}
  \right)^{1/2}\right] 
\end{equation*}
and $\Omega = \left(\omega^2 + \frac{c^2}{4}\right)^{1/2}$ coincides with a quantity denoted by $\omega'$ in \cite{meljanac05}. It is worth noting that in this case the operators $h$ and $O$ will have to be supplemented with some additional operators to provide us with an irreducible set of observables. Hamiltonians similar to (\ref{eq:h-conformal}) are currently encountered in a lot of problems, for instance in the important class of generalized Calogero models in $d$ dimensions (see, e.g., \cite{jakubsky, meljanac07}). So we do think that the correspondence of some non-Hermitian Hamiltonians of type (\ref{eq:gen-Swanson}) with them will prove rich in applications to many fields.\par
%
%--------------------------------------------------------------------------------------
%
In this paper, our key concern has been to propose a construction of the family of Hermitian Hamiltonians equivalent to the non-Hermitian oscillator one that would be independent of the realization of the su(1,1) generators making up such a Hamiltonian. As a by-product, this has opened the way towards new non-Hermitian Hamiltonians related by similarity transformations to Hermitian ones and therefore endowed with solid physical foundations.\par
%
%-----------------------------------------------------------------------------------------------------
%
As exemplified by the subtle discussion on p F79 of \cite{musumbu}, it should be clear that the detailed use of any of these non-Hermitian Hamiltonians may require a deeper and more careful analysis, which was beyond the scope of the present paper. All the technical conditions imposed on the metric are indeed important since we are working within infinite-dimensional Hilbert spaces, which demand special care (see, e.g., \cite{scholtz92, kret, mosta06}).\par
%
%---------------------------------------------------------------------------------------------------------
%
{}Finally, as far as the ambiguity of the metric is concerned, it is obvious that the conclusions of \cite{musumbu} apply here too. At this stage, we have no argument allowing us to elucidate this problem. Only experiments can tell us what is the appropriate choice in a given physical situation.\par
%
%=========================================================
%
\section*{Acknowledgments}

The author would like to thank H B Geyer and an anonymous referee for suggesting several amendments.
An interesting discussion with B Bagchi is also acknowledged.\par
%
%==============================================================
% 

\end{document}